\documentclass[twocolumn,pra,showpacs,superscriptaddress]{revtex4}
\usepackage{amssymb}
\usepackage{amsmath}
\usepackage{graphicx}
\usepackage{subfigure}
\usepackage{natbib}
\usepackage{epsfig}
\usepackage{amsfonts}
\usepackage{mathrsfs}
\usepackage{CJK}
\usepackage[toc,page,title,titletoc,header]{appendix}

\begin{document}
\title{Hierarchical environment-assisted dynamical speedup control}

\author{Kai Xu}
 \affiliation{Shandong Provincial Key
Laboratory of Laser Polarization and Information Technology,
School of Physics and Physical Engineering, Qufu Normal University, Qufu 273165, China}
 \affiliation{Beijing National Laboratory of Condensed Matter Physics, Institute of Physics,
Chinese Academy of Sciences, Beijing, 100190, China}

\author{Wei Han}
 \affiliation{Shandong Provincial Key
Laboratory of Laser Polarization and Information Technology,
School of Physics and Physical Engineering, Qufu Normal University, Qufu 273165, China}

\author{Ying-Jie Zhang}
\email{yingjiezhang@qfnu.edu.cn}
 \affiliation{Shandong Provincial Key
Laboratory of Laser Polarization and Information Technology,
School of Physics and Physical Engineering, Qufu Normal University, Qufu 273165, China}

\author{Yun-Jie Xia}
 \affiliation{Shandong Provincial Key
Laboratory of Laser Polarization and Information Technology,
School of Physics and Physical Engineering, Qufu Normal University, Qufu 273165, China}

\author{Heng Fan}
 \affiliation{Beijing National Laboratory of Condensed Matter Physics, Institute of Physics,
Chinese Academy of Sciences, Beijing, 100190, China}
\affiliation{Collaborative Innovation Center of Quantum Matter, Beijing, 100190, China}

\date{\today}
\begin{abstract}
We investigate the qubit in the hierarchical environment where the
first level is just one lossy cavity  while the second level is the
$N$-coupled lossy cavities. In the weak coupling regime between the
qubit and the first level environment, the dynamics crossovers from
the original Markovian to the new non-Markovian and from no-speedup
to speedup can be realized  by controlling the hierarchical
environment, i.e., manipulating the number of cavities or the
coupling strength between two nearest-neighbor cavities in the
second level environment. And we find that the coupling strength
between two nearest-neighbor cavities and the number of cavities in
the second level environment have the opposite effect on the
non-Markovian dynamics and speedup evolution of the qubit. In
addition, in the case of strong coupling between the qubit and the
first level environment,  we can be surprised to find that, compared
with the original non-Markovian dynamics, the added second level
environment cannot play a beneficial role on the speedup of the
dynamics of the system.
\end{abstract}

\maketitle

\section{\textbf{{Introduction}}}

With the rapid development of quantum information technology
\cite{Nielsen,Ladd}, the dynamics of open systems has drawn more and
more interest. Any open system inevitably takes environmental factor
considerations into account \cite{Petruccione}. The environment
interacting with the open system is divided into memory effect and
memoryless effect. In memoryless environment, the information flowed
in a single way, i.e., the information only flows from the system to
the environment results in  Markovian dynamics. While the
information  flows in both directions, i.e., the information can
backflow from the environment to the system and hence leads to the
non-Markovian dynamics in the memory environment. Recently, the
non-Markovian dynamics has been studied in theory
\cite{Anjh,Paz,Wolf,Tu,Breuer,Chruscinski,Xiong,Rivas,Znidaric,Zhangwm}
and in experiments \cite{Liubh,Madsen,Tang1} due to it plays a
leading role in many real physical processes such as quantum state
engineering, quantum control \cite{Xue,D'Arrigo9,Bylicka} and the
quantum information processing
\cite{Xiang,Bennett,Xu11,Aaronson,Duan}. And so far, the
non-Markovian dynamics have proven to be motivated by the strong
system-environment coupling, structure reservoirs, low temperatures,
initial system-environment correlations and so on. Furthermore, to
quantify non-Markovianity
\cite{Caruso,Huelga,Laine,Dajka,Smirne,Luo,Luxm,Chruscinski2,Lijg,Alimm,Piilo,Lorenzo,Plenio,Mat,Nguyen,Smirne2,Franco,Laine2,Liu4,Man1,Man2},
several measures \cite{Piilo,Lorenzo,Plenio} have been proposed and
the non-Markovian evolution process can absolutely be detected.

 More importantly, some efforts has been recently devoted to investigating the
 role played by the non-Markovianity
on the speed of evolution of  quantum system
\cite{Fan,Deffner,Sun,Meng,Zhangyj3,Xuzy,Cimmarusti,Anjh2,Zhangyj4}.
For the damped Jaynes-Cummings model, in which a qubit resonantly
interacts with a leaky single-mode cavity, reference \cite{Deffner}
has found that the non-Markovian effect could lead to speedup
dynamics process in the strong system-environment coupling regime.
And this theoretical result has been confirmed by increasing the
system-environment coupling strength and the controllable number of
atoms in the environment \cite{Cimmarusti}. So far, some works
usually consider the quantum system coupled to a single-layer
environment. However, in realistic scenarios, the quantum system
often inevitably couple to the multi-layer environments
\cite{Hanson1,Hanson2,Pla,Chekhovich,Tyryshkin}. Such as, the
neighbor nitrogen impurities constitute the principle bath for a
nitrogen-vacancy center, while the carbon-13 nuclear spins may also
couple to them \cite{Hanson1}. In a quantum dot the electron spin
may be affected strongly by the surrounding nuclei
\cite{Hanson2,Chekhovich}. A similar situation also occurs for a
single-donor electron spin in silicon \cite{Pla,Tyryshkin}.

Recently, motivated by these facts, many efforts have been devoted
to studying the effects of the multi-layer environments on the
non-Markovian dynamics of an open system. The dynamics of a
two-level atom has been studied in the presence of an overall
environment composed of two layers \cite{Nguyen}. In their model,
the first layer is just a single lossy cavity while the second layer
consists of a number of non-coupled lossy cavities. And the
non-Markovian dynamics of the atom is affected by the coupling
strength between the two layers and the number of cavities in the
second layer. But in the lab, some actual physical systems (such as
the couple cavity-array, the coupled spin chain and the coupled
superconducting resonators) usually have been chosen to implement
quantum computing and quantum simulation. So the influence of the
coupling between two nearest-neighbor cavities in the second layer
environment on the dynamics of the system can not be ignored.

So here we mainly study the dynamics of a qubit coupled with the two
layer environments. By using the quantum speed limit (QSL) time
\cite{Deffner,8,08} to define the speedup evolutional process, we
discuss that the influence of the coupling strength between two
nearest-neighbor cavities and the number of cavities in the
second-layer environment on the quantum evolutional speed of the
qubit. As we all known, in the weak coupling regime between the
qubit and the first-layer environment, the Markovian and no-speedup
dynamics of the qubit would be followed when the second-layer
environment has not been added. While in this paper, we elaborate
how the non-Markovian speedup dynamics of the system can be obtained
by controlling the coupling strength between two nearest-neighbor
cavities and the number of cavities in the second-layer environment.
Furthermore, we have also analyzed the influence of the parameters
of the added second-layer environment on the non-Markovian speedup
evolution of the qubit in the strong coupling regime between the
qubit and the first-layer environment.

\section{\textbf{Theoretical model}}

We consider that the total system consists of a qubit and the
hierarchical environment where the cavity $m_0$ and its
corresponding memoryless reservoir serve as the first-layer
environment for the two-level atom and the other coupled cavities
$m_{n}$ ($n=1$, 2, ..., $N$) and memoryless reservoirs involved act
as the second-layer environment, as depicted in Fig. 1.
\begin{figure}[tbh]
\includegraphics*[bb=68 198 575 505,width=7cm, clip]{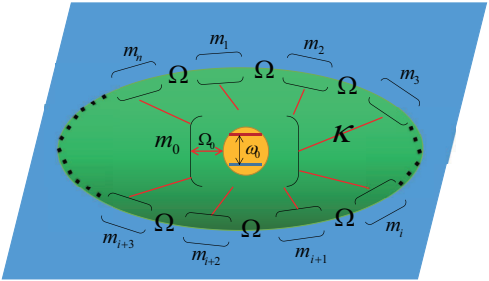}
\caption{Schematic representation of the model. The stratified lossy
cavities are selected as the controllable environment.}
\end{figure}
To be concrete, the qubit is coupled with strength $\Omega_{0}$ to a
mode $m_{0}$ which decays to a memoryless reservoir with a lossy
rate $\Gamma_{0}$ and then the mode $m_{0}$ is further coupled
simultaneously with strengths $\kappa$ to modes $m_{n}$ which also
decay to their respective memoryless reservoirs with rates
$\Gamma_{n}=\Gamma$. Besides, the coupling strength between two
nearest-neighbor cavities is $\Omega$ in the second-layer
environment. The total Hamiltonian can be given by $H=H_{0}+H_{I}$,
reads
\begin{eqnarray}
H_{0}&=&\frac{\omega_{0}}{2}\sigma_{z}+\omega_{c}a^{\dag}a+\sum_{n=1}^{N}\omega_{n}b^{\dag}_{n}b_{n},\nonumber\\
H_{I}&=&\Omega_{0}(\sigma_{+}a+\sigma_{-}a^{\dag})+\sum_{n=1}^{N}\kappa(ab^{\dag}_{n}+a^{\dag}b_{n})\nonumber\\
&+&\sum_{\langle
ij\rangle}\Omega(b^{\dag}_{i}b_{j}+b^{\dag}_{j}b_{i}). \label{01}
\end{eqnarray}
In the above expressions,
$\sigma_{z}=|1{\rangle}{\langle}1|-|0{\rangle}{\langle}0|$ is a
pauli operator for the qubit with transition frequency $\omega_{0}$,
$\sigma_{\pm}$ represent the raising and lowering operators of the
qubit, and $a$ $(a^{\dag})$ is the annihilation (creation) operator
of mode $m_{0}$ with frequency $\omega_{c}$. Besides, $b^{\dag}_{n}$
($b_{n}$) is the creation (annihilation) operator of mode $m_{n}$
($n=1$, 2, ..., $N$) with frequency $\omega_{n}$. $\langle
ij\rangle$ means the nearest neighbor cavities in the second-layer
environment. Then the density operator $\rho(t)$ of the total system
obeys the following master equation

\begin{eqnarray}
\frac{d\rho}{dt}=&-&i[H,\rho]-\frac{\Gamma_{0}}{2}(a^{\dag}a\rho-2a{\rho}a^{\dag}+{\rho}a^{\dag}a)\nonumber\\
&-&\sum_{n=1}^{N}\frac{\Gamma}{2}(b^{\dag}_{n}b_{n}\rho-2b_{n}{\rho}b^{\dag}_{n}+{\rho}b^{\dag}_{n}b_{n}) \label{02},
\end{eqnarray}

For simplicity, we suppose that the atom is initially in its excited
state $|1{\rangle}_{s}$, while all the cavities are in their ground
states $|0...0{\rangle}_{m_{0}....m_{n}}$, i.e., the initial state
of the total system is $\rho(0)=|10...0{\rangle}{\langle}10...0|$.
Since there exist at most one excitation in the total system at a
time, then $|\psi(t){\rangle}$ at time t can be written as
$|\psi(t){\rangle}=g(t)|10...0{\rangle}+c_{0}(t)|01...0{\rangle}+c_{1}(t)|001...0{\rangle}+...+c_{n}(t)|000...1{\rangle}$,
where $g(t)$, $c_{0}(t)$, $c_{1}(t)$, ..., $c_{n}(t)$ correspond to
probability amplitudes for the atom and modes $m_{0}$, $m_{1}$, ...,
$m_{n}$, respectively. Besides, the probability amplitudes $g(t)$,
$c_{0}(t)$, $c_{1}(t)$, ..., $c_{n}(t)$ of the system are governed
by the Schr$\ddot{o}$dinger equation, from which we can obtain,
\begin{eqnarray}
\begin{aligned}
i\dot{g}(t)&=\Omega_{0} c_{0}(t), \\
i\dot{c_{0}}(t)&=(-\frac{i}{2}\Gamma_{0})c_{0}(t)+\Omega_{0} g(t)+\sum_{n=1}^{N}\kappa c_{n}(t),\\
i\sum_{n=1}^{N}\dot{c_{n}}(t)&=\sum_{n=1}^{N}(2\Omega-\frac{i}{2}\Gamma)c_{n}(t)+\kappa N c_{0}(t). \label{04}
\end{aligned}
\end{eqnarray}
The solutions of the above equations can be obtained by means of
Laplace transform method. For convenience, we express the dynamics
of the qubit by the reduced density matrix in the system's bases
$\{|1{\rangle}_{s}$, $|0{\rangle}_{s}\}$ as
\begin{equation}
\rho(t)=\left(
  \begin{array}{cc}
    \rho_{11}(0)|g(t)|^{2} & \rho_{01}(0)g(t)^{\ast} \\
    \rho_{10}(0) g(t)      & \rho_{00}(0)+\rho_{11}(0)(1-|g(t)|^{2})\label{05} \\
  \end{array}
\right)
\end{equation}
where $\rho_{11}(0)=1$, $\rho_{00}(0)=\rho_{01}(0)=\rho_{10}(0)=0$.
Then in our scheme, it needs to be emphasized that we mainly focus
on the dynamical behavior of the atomic system can be modified by
the second-layer controllable environment in the weak qubit-$m_{0}$
coupling regime ($\Omega_{0}<\Gamma_{0}/4$) and the strong
qubit-$m_{0}$ couple regime ($\Omega_{0}>\Gamma_{0}/4$).

\section{\textbf{{Non-Markovian dynamics control}}}

In order to further illustrate the roles of the parameters in the
considered second-layer environment, i.e., the number $N$ of
cavities and the coupling strength $\Omega$ between two
nearest-neighbor cavities, in what follows we would describe how to
tune the controllable second-layer environment from Markovian to
non-Markovian by manipulating $N$ and $\Omega$. A general measure
$\mathbf{N}(\Phi)$ for non-Markovianity, which can be used to
distinguish the Markovian dynamics and the non-Markovian dynamics of
the quantum system, has been defined by Breuer $et$ $al$.
\cite{Breuer}.
 For a quantum process $\Phi(t)$,
$\rho^{s}(t)=\Phi(t)\rho^{s}(0)$, with $\rho^{s}(0)$ and
$\rho^{s}(t)$ denote the density operators at time $t=0$ and at any
time $t>0$ of the quantum system, respectively, then the
non-Markovianity $\mathbf{N}(\Phi)$ is quantified by
$\mathbf{N}(\Phi)=\max_{\rho^{s}_{1,2}(0)}\int_{\sigma>0}dt\sigma[t,\rho^{s}_{1,2}(0)],
$ with
$\sigma[t,\rho^{s}_{1,2}(0)]=\frac{d}{dt}\mathcal{D}(\rho^{s}_{1}(t),\rho^{s}_{2}(t))$
is the rate of change of the trace distance. The trace distance
$\mathcal{D}$ describing the distinguishability between the two
states is defined as \cite{Nielsen}
$\mathcal{D}(\rho^{s}_{1},\rho^{s}_{2})=\frac{1}{2}\|\rho^{s}_{1}-\rho^{s}_{2}\|,
$ where $\|M\|=\sqrt{M^{\dag}M}$ and $0{\leq}\mathcal{D}\leq1$. And
$\sigma[t,\rho^{s}_{1,2}(0)]\leq0$ corresponds to all dynamical
semigroups and all time-dependent Markovian processes. While a
process is non-Markovian if there exists a pair of initial states
and at certain time $t$ such that $\sigma[t,\rho^{s}_{1,2}(0)]>0$.
We should take the maximum over all initial states
$\rho^{s}_{1,2}(0)$ to calculate the non-Markovianity. In Refs.
\cite{Breuer,Lijg}, by drawing a sufficiently large sample of random
pairs of initial states, it is proved that the optimal state pair of
the initial states can be chosen as
$\rho^{s}_{1}(0)=(|0\rangle_{s}+|1\rangle_{s})/\sqrt{2}$ and
$\rho^{s}_{2}(0)=(|0\rangle_{s}-|1\rangle_{s})/\sqrt{2}$. Here, for
this optimal state pair, the rate of change of the trace distance
can be acquired
$\sigma[t,\rho^{s}_{1,2}(0)]=\partial_{t}|g(t)|^{2}$. Then the
non-Markovianity of the quantum system dynamics process from
$\rho^{s}(0)$ to $\rho^{s}(t)$ can be calculated by
$\mathbf{N}(\Phi)=\int^{t}_{0}(\partial_{t}|g(t)|^{2})|_{>0}dt$.

\begin{figure}[tbh]
\includegraphics*[bb=2 17 440 437,width=8cm, clip]{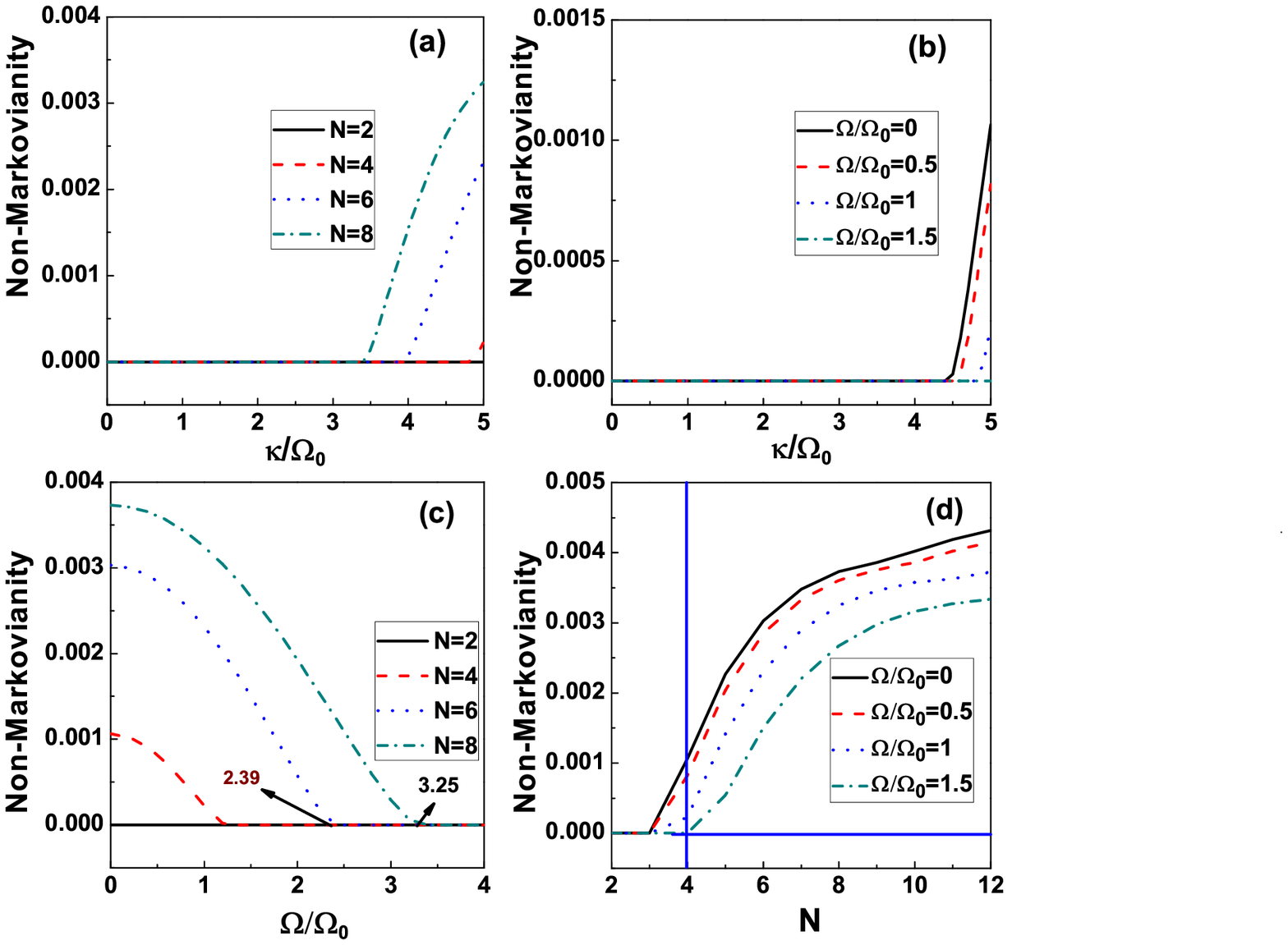}
\caption{(Color online) (a), (b) The non-Markovianity
$\mathbf{N}(\Phi)$ of the atomic system dynamics process from
$\rho^{s}_{0}=|1\rangle_{s}{\langle}1|$ to $\rho^{s}_{\tau}$ as a
function of the coupling strength $\kappa$ between the first-layer
environment and the second-layer environment in the weak
qubit-$m_{0}$ coupling regime $\Gamma_{0}=5\Omega_{0}$. (c) and (d)
The non-Markovianity $\mathbf{N}(\Phi)$ of the atomic system
dynamics process from $\rho^{s}_{0}=|1\rangle_{s}{\langle}1|$ to
$\rho^{s}_{\tau}$ as a function of the coupling strength $\Omega$
between two nearest-neighbor cavities and the number of cavities $N$
in the weak qubit-$m_{0}$ coupling regime $\Gamma_{0}=5\Omega_{0}$.
The other parameters are: (a) $\Omega=\Omega_{0}$,
$\Gamma=5\Omega_{0}$, $\tau=3$; (b) $N=4$, $\Gamma=5\Omega_{0}$,
$\tau=3$; (c) and (d) $\kappa=5\Omega_{0}$, $\Gamma=5\Omega_{0}$,
$\tau=3$.}
\end{figure}

As we all know, if there are no any other-layer environment, the
system's dynamics mainly depends on the parameters $\Omega_{0}$ and
$\Gamma_{0}$ in such a way that $\Gamma_{0}>4\Omega_{0}$
($\Gamma_{0}<4\Omega_{0}$), identified as the weak (strong) coupling
regime, leads to Markovian (non-Markovian) dynamics. Then in the
case of adding the second-layer environment, the non-Markovianity
$\mathbf{N}(\Phi)$ of the atomic dynamics process in the weak
qubit-$m_{0}$ coupling regime from $\rho^{s}_{0}$ to
$\rho^{s}_{\tau}$ as function of the controllable second-layer
environment parameters ($\kappa$, $\Omega$, $N$) with the actual
evolution time $\tau=3$, has been plotted in Fig. 2. Firstly, by
fixing $\Omega=\Omega_{0}$, $\Gamma_{0}=5\Omega_{0}$ in the Fig.
$2(a)$, in the case $N=2$, the non-Markovian dynamics of the atomic
system cannot be obtained by tuning the $m_{0}-m_{n}$ coupling
strength $\kappa$. However, if the lossy cavity $m_{0}$
simultaneously interacts the second-layer environment constituted by
more than the number of cavities $N=2$, then a remarkable dynamical
crossover from Markovian behavior to non-Markovian behavior can
occur at a certain critical coupling strength $\kappa_{c1}$. When
$\kappa<\kappa_{c1}$, the dynamics process abides by Markovian
behavior, and then the non-Markovianity increase monotonically with
increasing $\kappa$. Similarly, in the case $\Omega=1.5\Omega_{0}$
in the Fig. $2(b)$, no matter how to adjust the parameter $\kappa$,
the dynamics process is Markovian. However, the non-Markovian
dynamics can be triggered at a threshold $\kappa_{c2}$  by
decreasing the coupling strength $\Omega$ between two
nearest-neighbor cavities, such as, $\Omega=0, 0.5\Omega_{0},
\Omega_{0}$. And it is clear to point that, when
$\kappa<\kappa_{c2}$ the dynamics process is Markovian, and the
non-Markovianity increases with increasing $\kappa$ in the case
$\kappa>\kappa_{c2}$. Finally, in the weak qubit-$m_{0}$ coupling
regime, it needs to be emphasized that, the larger the value
$\kappa$ ($N$), the smaller value $\Omega$ should be requested to
trigger the non-Markovianity of the system.

In the following, in order to more intuitively explain the effect of
parameters $\Omega$ and $N$ on the non-Markovianity of the system in
weak qubit-$m_{0}$ coupling regime, we firstly fix
$\kappa=5\Omega_{0}$ larger than the above critical value
$\kappa_{c1}$ or $\kappa_{c2}$ in the Figs. $2(c)$ and $2(d)$. It is
worth noting that, the dynamical crossover from Markovian behavior
to non-Markovian behavior for the atomic dynamics process
($\rho^{s}_{0}$ to $\rho^{s}_{\tau}$) could appear by tuning
$\Omega$, $N$. When the value $N$ is confirmed in Fig. 2(c), in the
case $\Omega>\Omega_{c}$ ($\Omega_{c}$ means the critical value of
$\Omega$), the dynamics process always abides by Markovian behavior,
but the non-Markovianity increases monotonically with reducing
$\Omega$ when $\Omega<\Omega_{c}$. That means the increasing of the
coupling strength $\Omega$ in the second-layer environment could
suppress the non-Markovian dynamics of the atomic system. Moreover,
the larger the value $N$, the larger the critical value $\Omega_{c}$
should be acquired. Take the cases in Fig. $2(c)$: when $N=6$, we
find the critical value $\Omega_{c1}=2.39\Omega_{0}$. While in the
case $N=8$, $\Omega_{c2}=3.25\Omega_{0}$ should be given.
Furthermore, it is interesting to find that, by fixing
$\Omega=1.5\Omega_{0}$ in the Fig. $2(d)$, the dynamics process
abides by Markovian up to $N=3$, but becomes non-Markovian starting
from $N=4$. However, the non-Markovian dynamics is induced from
$N=3$ by decreasing the coupling strength $\Omega$, say, $\Omega=0,
0.5\Omega_{0}, \Omega_{0}$. In general, it is worth noting that a
larger $\Omega$ ($N$) could lead to a larger
critical $N$ ($\Omega$). Finally, in the weak qubit-$m_{0}$ coupling
regime, we find that the non-Markovian dynamic can be triggered by
manipulating the added second-layer environment parameters.
\begin{figure}[!tbh]
\includegraphics*[bb=0 106 545 359,width=8cm, clip]{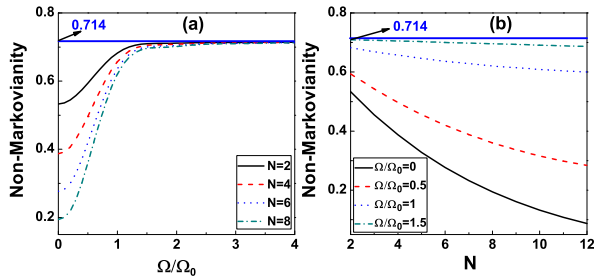}
\caption{(Color online) (a), (b) The non-Markovianity
$\mathbf{N}(\Phi)$ of the atomic system dynamics process from
$\rho^{s}_{0}=|1\rangle_{s}{\langle}1|$ to $\rho^{s}_{\tau}$ as a
function of the coupling strength $\Omega$ between two
nearest-neighbor cavities and the number of cavities $N$ in the
strong qubit-$m_{0}$ coupling regime $\Gamma_{0}=0.2\Omega_{0}$. The
parameters are: $\Gamma=0.2\Omega_{0}$, $\kappa=0.2\Omega_{0}$,
$\tau=3$.}
\end{figure}

It is well known that in the absence of the second-layer
environment, the dynamics process of the system is non-Markovian in
the strong qubit-$m_{0}$ coupling regime. Next, Figs. $3(a)$ and
$3(b)$ show the non-Markovianity $\mathbf{N}(\Phi)$ of the atomic
dynamics process from $\rho^{s}_{0}$ to $\rho^{s}_{\tau}$ is
affected by the parameters $\Omega$ or $N$ with the actual evolution
time $\tau=3$ when the strong qubit-$m_{0}$ coupling satisfied
$\Gamma_{0}=0.2\Omega_{0}$. The blue solid line $N(\Phi)=0.714$ in
Fig. 3 means the value of non-Markovianity without the second-layer
environment. By fixing $N$ in Fig. 3(a), the non-Markovianity
increases and tends to the original non-Markovianity $N(\Phi)=0.714$
of the atomic system with increasing the coupling strength $\Omega$
between two nearest-neighbor cavities, implying the coupling
strength $\Omega$ between the two nearest-neighbors in the
second-level environment can promote the non-Markovian dynamics of
the system. Furthermore, by fixing $\Omega$ in the Fig. 3(b), the
non-Markovianity of the system increases with decreasing the number
of cavities $N$ in the second-layer environment. And it is clear to
find that, irrespective of $\Omega$, $N$, the non-Markovianity is
always below the original non-Markovianity ($0.714$) of the system.
That is to say, compared with the previous non-Markovian dynamics in
the strong qubit-$m_{0}$ coupling regime, the added second-layer
environment can not contribute to promote the non-Markovianity of
the system. Besides, in stark contrast, when the parameters $N$ and
$\Omega$ make it possible to trigger the non-Markovianity of the
system under the weak qubit-$m_{0}$ coupling regime, the
non-Markovianity of the system is weakened under the strong
qubit-$m_{0}$ coupling regime at this time. That is to say, in the
cases of weak qubit-$m_{0}$ coupling regime and strong qubit-$m_{0}$
coupling regime, the second-layer environment parameters $\Omega$
and $N$ have different effects on the non-Markovian dynamics of the
system. This is a newly noticed phenomenon.

\section{\textbf{{Quantum speedup of the atomic dynamics}}}

Recently, the relationship between non-Markovianity and the QSL
times has been given by  $\frac{\tau_{QSL}}{\tau}
=\frac{1-|g(t)|^{2}}{2\mathbf{N}(\Phi)+1-|g(t)|^{2}}$
\cite{Xuzy,Zhangyj4}. So the same critical values ($\Omega$, $N$)
can be obtained from the Markovian process to the non-Markovian
process and from no-speedup of quantum evolution to speedup. And the
above equation implies that the stronger the non-Markovianity, the
lower the QSL times (the quantum speedup evolution would occur).
Then in our investigated model, it is easily to find that the
purpose of accelerating the evolution of the quantum system can also
be achieved by the controllable non-Markovianity discussed aboved.
Obviously, in this section, we mainly focus on how the  coupling
strength $\Omega$  and the number of cavities $N$ to accelerate the
evolution of quantum system. In order to characterize how fast the
quantum system evolves, here we use the definition of the QSL time
for an open quantum system, which can be helpful to analyze the
maximal speed of evolution of an open system. The QSL time between
an initial state $\rho^{s}(0)=|\phi_{0}\rangle\langle\phi_{0}|$ and
its target state $\rho^{s}(\tau)$ for open system is defined by
\cite{Deffner}
$\tau_{QSL}=\sin^{2}[\mathbf{B}(\rho^{s}(0),\rho^{s}(\tau))]/\Lambda^{\infty}_{\tau}$,
where
$\mathbf{B}(\rho^{s}(0),\rho^{s}(\tau))=\arccos\sqrt{\langle\phi_{0}|\rho^{s}(\tau)|\phi_{0}\rangle}$
denotes the Bures angle between the initial and target states of the
system, and
$\Lambda^{\infty}_{\tau}=\tau^{-1}\int^{\tau}_{0}\|\dot{\rho}^{s}(t)\|_{\infty}dt$
with the operator norm $\|\dot{\rho}^{s}(t)\|_{\infty}$ equaling to
the largest singular value of $\dot{\rho}^{s}(t)$. When the ratio
between the QSL time and the actual evolution time equals one, i.e.,
$\tau_{QSL}/\tau=1$, the quantum system evolution is already along
the fastest path and possesses no potential capacity for further
quantum speedup. While for the case $\tau_{QSL}/\tau<1$, the speedup
evolution of the quantum system may occur and the much shorter
$\tau_{QSL}/\tau$, the greater the capacity for potential speedup
will be.

\begin{figure}[tbh]
\includegraphics*[bb=50 2 439 461,width=7.5cm, clip]{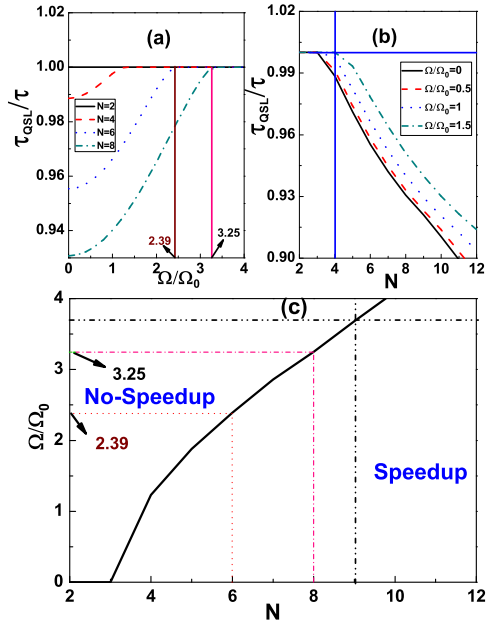}
\caption{(Color online) (a), (b) The QSL time for the atomic system
dynamic process with $\Gamma_{0}=5\Omega_{0}$ in the weak
qubit-$m_{0}$ coupling regime as a function of the coupling strength
$\Omega$ and the number of cavities $N$ in the second-layer
environment. (c)Phase diagram of QSL time in the $\Omega$-$N$ plane
with $\Gamma_{0}=5\Omega_{0}$ in the weak qubit-$m_{0}$ coupling
regime. The parameters are: $\Gamma=5\Omega_{0}$,
$\kappa=5\Omega_{0}$.}
\end{figure}

Then, for the weak qubit-$m_{0}$ coupling regime, the variations of
the QSL time $\tau_{QSL}/\tau$ with respect to $\Omega$ and $N$ are
plotted in Fig. 4. It is clearly found that, in the Fig. $4(a)$, the
quantum system has no potential capacity for the further quantum
speedup until the number of cavities in the second-layer environment
above $N=2$. In order to verify, in the case $N=2$ in Fig. $4(a)$,
the no-speedup evolution ($\tau_{QSL}/\tau=1$) could be followed,
and the speedup evolution ($\tau_{QSL}/\tau<1$) would occur when the
coupling strength $\Omega$ is less than a certain critical coupling
strength $\Omega_{c}$ in the cases $N=4$, $6$, $8$. In addition, it
is worth noting that, the larger the value $N$, the larger the
critical value $\Omega_{c}$ may be requested. As shown in the Fig.
$4(a)$, when $N=6$, $\Omega_{c1}=2.39\Omega_{0}$. While $N=8$,
$\Omega_{c2}=3.25\Omega_{0}$. That is same as the above discussion
for the non-Markovianity. As for Fig. 4(b), by fixing
$\Omega=1.5\Omega_{0}$, the speedup evolution may occur when the
number of cavities $N>4$ in the second-layer environment. However,
by considering $\Omega=0$, $0.5\Omega_{0}$, $\Omega_{0}$, the speedup
evolution can be induced by $N>3$. So we get an interesting
conclusion that the coupling strength $\Omega$ and the number of
cavities $N$ in the second-layer environment have the opposite
effect on the speedup quantum evolution of the atomic system,
namely, the quantum evolution would speedup by decreasing $\Omega$
or increasing $N$. So in the weak qubit-$m_{0}$ coupling regime, the
purpose of accelerating evolution can be achieved by controlling the
hierarchical environment.

In order to clear the region of parameters in which the speedup
dynamics of the system can be eventuated in the weak qubit-$m_{0}$
coupling regime, Fig. $4(c)$ describes the $\Omega-N$ phase
diagrams. And the transition points from no-speedup to speedup
regime could be acquired. Take the cases in Fig. $4(c)$: when the
value $N=6$, $8$ are fixed, the corresponding critical values of
$\Omega$ have also been respectively given by
$\Omega_{c1}=2.39\Omega_{0}$ and $\Omega_{c2}=3.25\Omega_{0}$. Still
further, by fixing $N$, when $\Omega<\Omega_{c}$, the quantum
speedup evolution would occur and then remain no-speedup with the
increasing $\Omega$. It is worth noting that the number of the
cavities in the second-layer environment must be more than $2$. The
smaller $\Omega$ and the larger $N$ should be manipulated to drive
the purpose of speedup evolution of the quantum system in the weak
qubit-$m_{0}$ coupling regime.

For the strong qubit-$m_{0}$ coupling regime, by fixing
$\Gamma_{0}=0.2\Omega_{0}$ and $\kappa=0.2\Omega_{0}$, we plot the
variations of the QSL time $\tau_{QSL}/\tau$ with respect to
$\Omega$ and $N$ in Fig. 5. Then this blue solid line indicates that
when no adding the second-layer environment, the ratio value of
$\tau_{QSL}/\tau$ is equal to $0.166$. Clearly shown in Fig. 5, the
value of $\tau_{QSL}/\tau$ decreases and tends to 0.166 as the
increasing of $\Omega$ in the Fig. $5(a)$. That is to say, the
increasing of $\Omega$ can accelerate the evolution of the system in
our two-layer environments. Besides, by fixing $\Omega$ in the Fig.
$5(b)$, the QSL time always increase as the increasing of the number
of cavities $N$. However, regardless of how to tune parameters $N$
or $\Omega$ in the Fig. 5, the QSL time is always larger than 0.166.
That is to say, although $N$ or $\Omega$ can affect the QSL time of
the dynamics process of the system in the strong qubit-$m_{0}$
coupling regime, the added second level environment cannot play a
beneficial role on the speedup of the dynamics of the system.

\begin{figure}[tbh]
\includegraphics*[bb=0 98 534 389,width=8cm, clip]{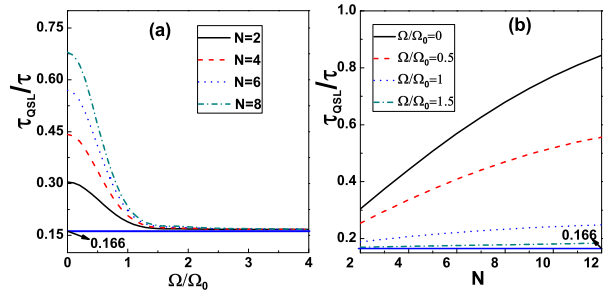}
\caption{(Color online) (a), (b) The QSL time for the atomic system
dynamic process with $\Gamma_{0}=0.2\Omega_{0}$ in the strong
qubit-$m_{0}$ coupling regime as a function of the coulping-strength
$\Omega$ and the number of cavities $N$.  The parameters are:
$\Gamma=0.2\Omega_{0}$, $\kappa=0.2\Omega_{0}$.}
\end{figure}

\section{\textbf{{Conclusion}}}

In conclusion, we have investigated the dynamics of the qubit in a
controllable hierarchical environment where the first-layer
environment is a lossy cavity $m_{0}$ and the second-layer
environment are the other coupled lossy cavities. Some interesting
phenomena are observed. On the one hand, by controlling the number
of cavities $N$ and the coupling strength $\Omega$ between two
nearest-neighbor cavities in the second-layer, two dynamical
crossovers of the quantum system, from the original Markovian to the
new non-Markovian dynamics and from no-speedup evolution to speed
evolution, have been achieved in the weak qubit-$m_{0}$ coupling
regime. And it is worth noting that the coupling strength $\Omega$
and the number of cavities $N$ have the opposite effect on the
non-Markovian dynamics and speedup evolution of the atomic system in
the weak qubit-$m_{0}$ coupling regime. Furthermore, the transitions
from no-speedup phase to speedup phase and from Markovian to
non-Markovian effect for the system, have been demonstrated in our
work. On the other hand, by considering the strong qubit-$m_{0}$
coupling regime, compared with the original non-Markovian dynamics
without the second-layer environment, the added second-layer
environment can not promote the speedup evolution of the system. And
for the weak qubit-$m_{0}$ coupling regime and strong qubit-$m_{0}$
coupling regime, the second-layer environment parameters $\Omega$
and $N$ have different effects on the non-Markovian speedup dynamics
of the system. Our hierarchical environment-assisted non-Markovian
speedup dynamics control is essential to most purposes of quantum
optimal control.

\section{\textbf{{Acknowledgements}}}
This work is supported by the National Natural Science Foundation of
China (11647171, 61675115, 91536108), MOST of China (2016YFA0302104,
2016YFA0300600), and the Foundation of Chinese Academy of Sciences
(XDB01010000, XDB21030300).

\end{document}